\documentclass[prl,twocolumn,superscriptaddress,floatfix,citesort]{revtex4}

\usepackage{graphicx, epsfig}
\usepackage{amsmath, amssymb}
\usepackage{color}
\usepackage{bbm}
\usepackage{subfigure}
\usepackage{fancyhdr}
\newcommand{\bra}[1]{\langle #1 |}
\newcommand{\ket}[1]{| #1 \rangle}

\newcommand{\comment}[1]{\textcolor{red}{\textit{#1}}}
\renewcommand{\comment}[1]{}                        
\let\note\comment
\newcommand{\SKIP}[1]{}
\newcommand{\citen}[1]{\note{\textcolor{black}{\bf #1}}\cite{#1}}

\newcommand{\myheading}[1]{\subsection{#1}}
\newcommand{\myheadingA}[1]{\subsubsection{#1}}
\newcommand{\myheadingB}[1]{\textit{#1}}
\newcommand{\Fig}[1]{Fig.~\ref{#1}}
\newcommand{\bit}{\begin{itemize} \setlength{\itemsep}{0ex} \setlength{\topsep}{0ex} } 
\newcommand{\eit}{\end{itemize}}
\newcommand{\be}{\begin{equation}}
\newcommand{\ee}{\end{equation}}
\newcommand{\bea}{\begin{eqnarray}}
\newcommand{\eea}{\end{eqnarray}}
\newcommand{\half}{\frac{1}{2}}
\let\gsim=\gtrsim

\let\mytitle=\title
\let\myauthor=\author
\let\myaffiliation=\affiliation

\begin{document}

\title{Quantum Bowling: Particle-hole transmutation in one-dimensional strongly interacting lattice models}
\author{Martin Ganahl}
\affiliation{Institut f. Theoretische Physik, Technische Universit\"at Graz, Petersgasse 16, 8010 Graz, Austria}
\author{Masud Haque}
\affiliation{Max-Planck-Institut f\"ur Physik komplexer Systeme, N\"othnitzer Stra\ss e 38, 01187 Dresden, Germany}
\author{H.G. Evertz}
\email{evertz@tugraz.at}
\affiliation{Institut f. Theoretische Physik, Technische Universit\"at Graz, Petersgasse 16, 8010 Graz, Austria}
\date{February 11, 2013} 
\pagestyle{fancy}
\fancyhead[L]{Ganahl, Haque, Evertz}
\fancyhead[R]{Quantum Bowling}
\fancyfoot[C]{\thepage}

\begin{abstract}
We study the scattering of a soliton-like propagating particle with a wall of  
bound particles, in several strongly interacting one-dimensional lattice models with discrete degrees of freedom.
We consider spin-polarized fermions (anisotropic Heisenberg spin chain), 
the fermionic Hubbard model, and the Bose Hubbard model,
using precise numerical time dependent Density Matrix Renormalization Group techniques.
We show that in all integrable models studied, there is no reflection.
Instead, an incoming particle experiences particle-hole transmutation upon entry and exit of the wall,
and travels inside the wall as a hole, analoguous to Klein tunneling, even though the dispersion is highly nonlinear
and there is no external potential.
{\em Two} particles are added to the wall on the incoming side and removed on the opposite side.
For spin-polarized fermions a single transmitted particle thus shifts the wall by two lattice sites,
in complete contrast to classical physics.
For both Hubbard models, the wall shifts by one doubly occupied single site.
In the nonintegrable models studied, the same process occurs in linear superposition with backscattering events.
We demonstrate a corresponding fermionic quantum Newton's cradle
and a metamaterial with ``tachyonic'' modes travelling faster than in an empty system.
We present a possible atomic scale signal counter for spintronics. 
 Our scenario should be realizable in future cold atom experiments.
\end{abstract}


\maketitle

The investigation of time evolution in non-equilibrium situations is a fast-expanding
frontier in quantum many-particle physics.  With the development
of relevant experimental techniques, 
e.g.\  in cold-atom setups \citen{Cold-atom-review,Kinoshita-cradle}, 
situations that were quite academic a decade or two ago have become accessible.
The study of
coherent non-dissipative dynamics far from equilibrium has received a
particular boost.

\begin{figure}
\includegraphics[width=\columnwidth]{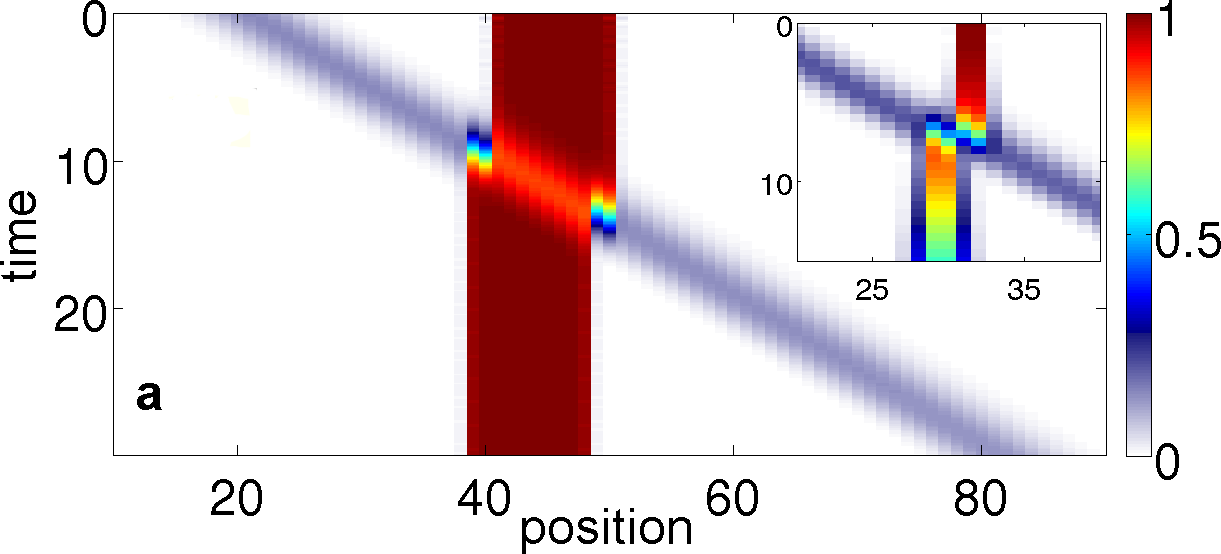}
\vskip5ex 
\includegraphics[width=\columnwidth]{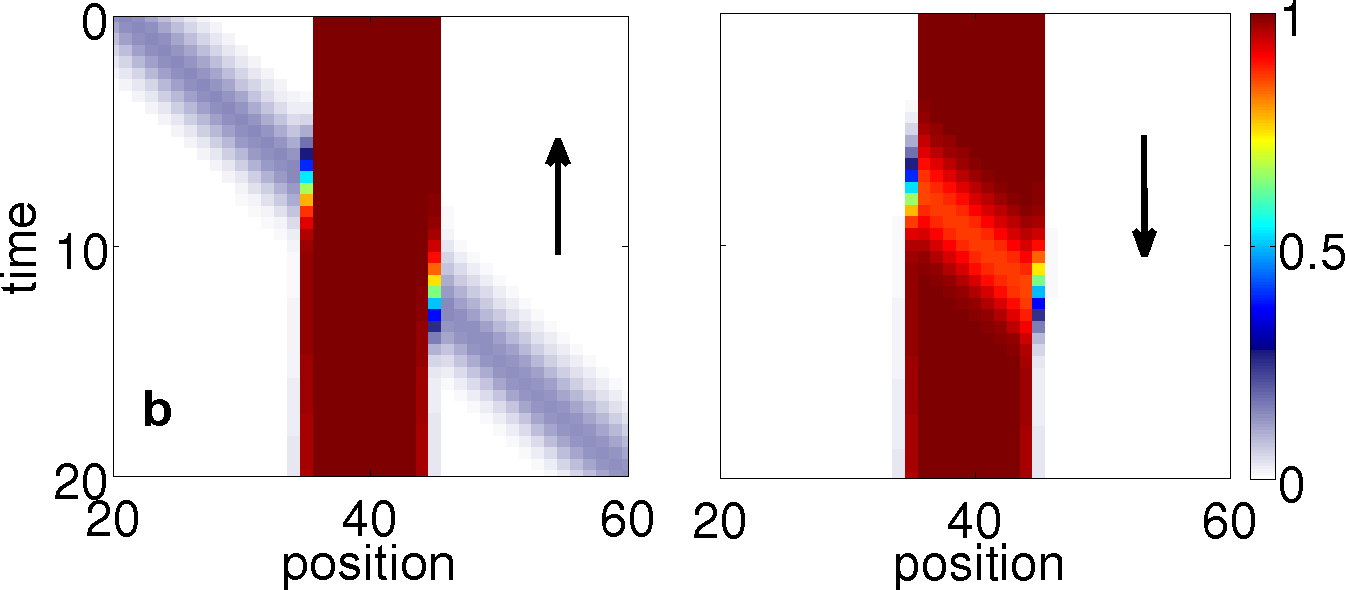}
\vskip5ex
\includegraphics[width=\columnwidth]{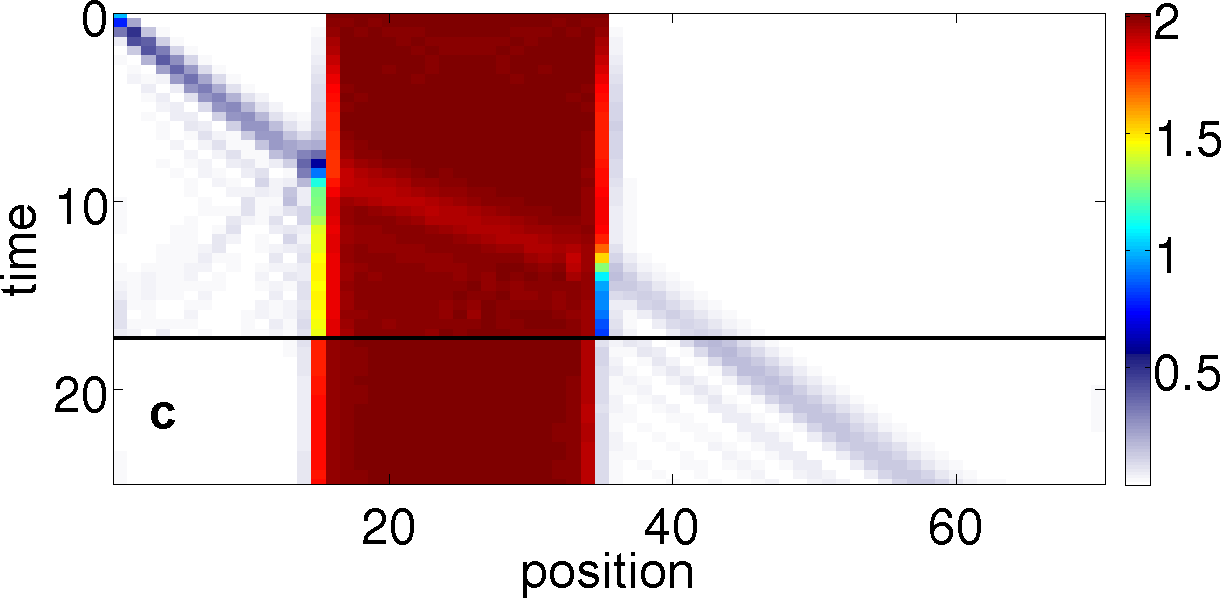}

\caption{Quantum bowling: an incoming single particle in Gaussian shape
        crosses a wall of bound particles on $N$ consecutive sites.
        It undergoes particle-hole transmutation and shifts the wall by 2 particles.
        The figures show particle density $\langle n_i(t)\rangle$.
(a) Spin-polarized fermions with $V/t=10$. The 10-site wall is moved by 2 lattice sites. 
    Inset: A wall of just 2 sites is still moved ($V/t=20$). 
(b) Fermi Hubbard model. Left: up-spin density, right: down-spin density.
    Wall of $N=10$ doubly occupied sites, $U/t=100$.
(c) Bose Hubbard model. Wall of $N=18$ doubly occupied sites, $U/t=30$.
    The incoming particle is magnon-like \citen{supplement}.
    Lower part in (c): density on condition that a particle is present to the right of the wall.}
\label{fig:ThreeModels}
\end{figure}

An emerging theme is the dynamics and interaction of excitations over simple
states.  For condensed-matter models known and intensively studied for many
decades, surprisingly little is known about the dynamical evolution of
excitations that are simple but are not eigenstates. Obvious examples are the
dynamics of a few interacting particles on lattices, or a few flipped spins in
an otherwise ferromagnetic background.  In Hubbard models, wide
interactions ``repulsively bind'' on-site pairs and
other clusters or combinations of bosons \citen{repulsive-binding-Winkler}.  
In models with nearest neighbor couplings like spin chains, bound states (``multi-magnons'')
exist even at small couplings and their non-equilibrium dynamics 
has been the subject of recent research
\citen{Haque-edge-XXZ,Ganahl11,WoellertHonecker-XXZsoliton,MosselCaux}. 
Interacting models and dynamics with single site resolution are now within experimental reach \citen{SingleSite}.

Collision processes are fundamental for
analysing the physical laws governing the dynamics 
of classical and quantum systems. 
A famous classical realization is Newton's cradle,
exhibiting the effects of energy and momentum conservation.
A quantum version has been realized in cold atom experiments \citen{Kinoshita-cradle}.

In this paper we analyse scattering in one-dimensional strongly interacting quantum mechanical lattice models
with discrete degrees of freedom
using standard precise techniques for coherent many body time evolution \citen{tebd-rem,supplement}.
We find intriguing phenomena 
caused by the discrete quantum nature of the system. 
We consider three prominent models, spin-polarized fermions, the fermionic Hubbard model, 
and the Bose Hubbard model.
We will first briefly describe the main phenomena, exhibited in \Fig{fig:ThreeModels}.
Then we analyse the physics behind the observed behavior, arguing that 
it follows from conservation laws and the discreteness of the models. 
Finally we discuss a fermionic quantum Newton's cradle, a metamaterial with a tachyonic mode, and other possible
 applications. 
The supplementary material contains details of the calculations and further examples.

\myheading{Main phenomena.}
The setup of what we call {\em quantum bowling} consists of an almost stationary  ``wall'' of particles
sitting on consecutive sites,
on top of an empty lattice.
The wall is hit by a single 
soliton-like
particle (\Fig{fig:ThreeModels}).  
The ensuing dynamics are qualitatively independent of details of the initial state or magnitude of the couplings, 
within a large range. 
For integrable models \citen{Takahashi-book,integrable-scattering}
(\Fig{fig:ThreeModels}a,b), {\em no backscattering} occurs.
Instead, the incoming particle 
is transmitted as a {\em hole} through the wall,
i.e.\ there is {   particle hole transmutation}.
This situation resembles a manybody version of Klein tunneling \citen{KleinTunneling},
even though the dispersion of contributing modes is strongly non-linear and there is no external potential.
Strikingly, because of particle number conservation,
there are then {\em two} particles that remain stuck at the front of the wall, 
instead of the familiar classical result as in Newton's cradle,
where just the single incoming particle would be added to the wall.
When the transmitted hole exits the wall, it is converted into a particle again,
so that two particles vanish from the right side of the wall.
For spin-polarized fermions (\Fig{fig:ThreeModels}a) 
a single incoming particle {\em shifts the whole wall by two sites} to the left,
in contrast to the classically expected shift by a single site.
The exiting particle itself is shifted to the right.
At large coupling, the shift is by two sites 
and the shape of the wave packet is unchanged by the transmission.
In both Hubbard models (\Fig{fig:ThreeModels}b,c), the wall consists of doubly occupied sites,
and it is shifted by two particles, namely one doubly-occupied site.
Again there is particle hole transmutation.
In the Fermi-Hubbard model (\Fig{fig:ThreeModels}b), an incoming up-spin fermion
becomes a down-spin hole inside the wall.
In non-integrable models like the Bose Hubbard model (\Fig{fig:ThreeModels}c),
there is a finite probability for backscattering.
However, the final state is essentially a linear superposition of 
(i) a backscattering event and (ii) transmission with particle-hole transmutation and wall-shift
as described above.
Indeed, at large couplings there is very little entanglement between a transmitted particle 
and the shifted wall, even when the transmitted particle itself is in an entangled Gaussian state.
In the following, we will discuss the models individually.

\myheading{Spin-polarized fermions.}
%
The dynamics of spin-polarized fermions is governed by the Hamiltonian 
\begin{equation}\label{eq:tV}
  H_{tV} =  t  \sum_{i} \left( c^{\dag}_i c_{i+1} +  c^{\dag}_{i+1} c_{i} \right) 
           +V  \sum_{i} (n_i-\frac{1}{2})(n_{i+1}-\frac{1}{2}) 
\end{equation}
where $t$ and $V/t$ parametrize the kinetic and interaction energy of the fermions on the lattice, respectively,
$c_i^\dag$ and $c_i$ are creation and annihilation operators at site $i$, and $n_i=c^\dag_i c_i$.
This model is equivalent by a Jordan-Wigner transformation  
to the 1D spin $\half$ Heisenberg XXZ model
with anisotropy $\Delta=J_z/J_x = V/2t$.
We use $t=1$ as the unit of energy and inverse time for all models, 
and discuss $V>2t$. 
We note that when starting from a product state,
results are identical for either sign of $U$ \citen{Schneider12}, 
and, by adapting the arguments in ref.~\citen{Schneider12}, for either sign of $V/t$,
i.e.\ for attractive and repulsive models.

The model is exactly solvable by Bethe ansatz. 
It contains eigenstates made up of strings of $M$ 
bound particles in an otherwise empty lattice \citen{MosselCaux,Ganahl11,Haque-edge-XXZ},
which become more compact as $V/t$ increases.
Their energy is \citen{Takahashi-book} 
\begin{equation}
 E = 2t ( \cosh(M\phi) - \cos K ) {\sinh\phi}/{\sinh(M\phi)}    
\end{equation}
where $\cosh \phi = V/2t$ and $K$ is crystal momentum.
Their time evolution is dominated by \citen{Ganahl11}
a maximum velocity 
$2t\, {\sinh\phi}/{\sinh(M\phi)} $ 
which decreases like $\frac{2t}{(V/t)^{M-1}}$ for $V\gg 2t$.
At large $V/t$ and large $M$, this velocity becomes exponentially small;
M-strings are therefore stable on a very long time scale even when prepared at a fixed position in space.

In the initial state we prepare a ``wall'', 
a product state 
of $N=10$ consecutive particles on an empty lattice. This is not an eigenstate,
but it is instead made up of all subdivisions of $N$ into M-strings.
Since smaller strings have larger velocities, they will in time ``evaporate'' \citen{Heidr}
from the wall.
At large $V/t$ the dominant contribution \citen{MosselCaux} 
is an almost stationary $M=N$ string, followed by the subdivision into $M=9$ 
and a single fermion with maximum velocity $2t$ \citen{supplement}.
In the initial state we also prepare a single particle in a Gaussian superposition \citen{supplement} 
of width $4$ and momentum $-\pi/2$ 
(or alternatively as a local particle at the left boundary \citen{supplement})
which moves towards the wall with velocity $2$. 

Being integrable, the $tV$ model 
contains a macroscopic number of conserved quantities,
of which particle number $n=\sum_l n_l$, energy $E$, and
energy current 
(thermal current) 
$j^E = \sum_l j^E_l$ are especially relevant, with 
\begin{align}
 j_l^E\;=\;&\nonumber    itV ( c^{\dagger}_{l-1} c_l  (n_{l+1}-\frac{1}{2}) + (n_{l-1}-\frac{1}{2}) c^{\dagger}_l c_{l+1} )\\ \nonumber
                        +\; &it^2  c^{\dagger}_{l-1} c_{l+1}
                   \; \; + \; \; h.c. \; . 
\end{align}
Since the interaction is local, the energy current is conserved separately in spatially disconnected regions.

  \begin{figure}[t]
      \includegraphics[width=\columnwidth]{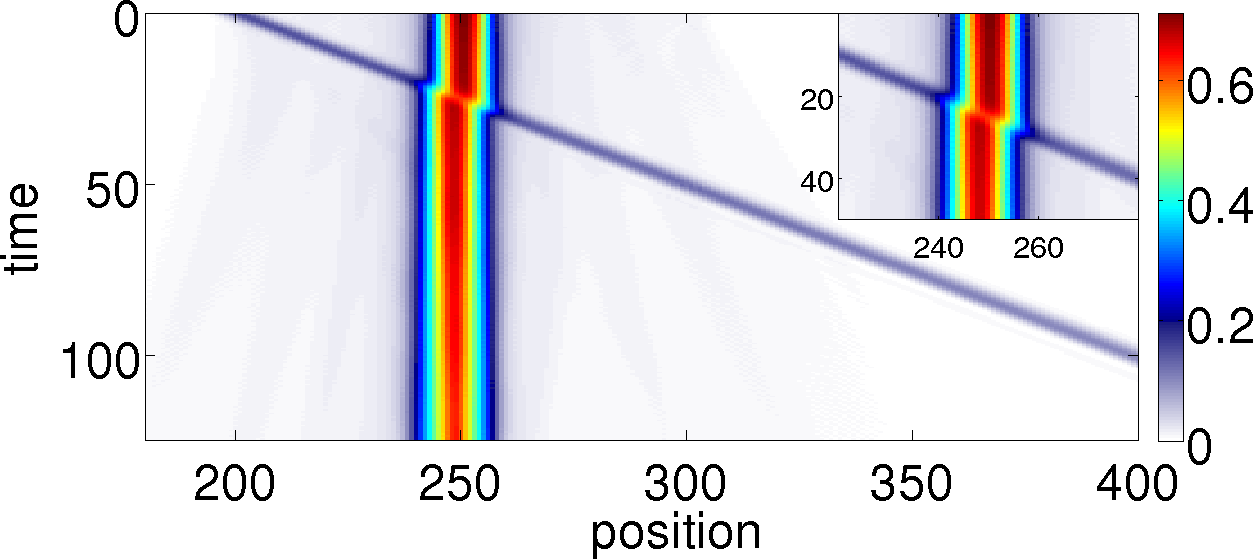}
    \caption{Small anisotropy: density of spin-polarized fermions at $V/t=2.2$ ($\Delta=1.1$ 
         in the equivalent Heisenberg model); evaporatively cooled initial state.
      Inset: scattering region at larger scale.
    } 
    \label{fig:smallDelta}
  \end{figure}

The initial state has a finite right-moving energy current, localized around the incoming particle
\citen{supplement}.
This current is conserved after the particle contacts the wall.
Yet the wall is ``full'' and cannot accomodate an additional particle.
Only a {\em hole} can move through the wall.
Inside the wall the hole possesses the same energy current and the same energy as the incoming particle on the empty lattice 
because of particle-hole symmetry.
However, because of particle number conservation, a particle-hole {\em pair} must have been created,
and there are now {\em two} particles located to the left of the wall.
Because of current conservation and energy conservation,
they cannot move: 
if there was a backscattered particle, it would have additional energy and a left moving energy current,
which would have to be compensated by a right moving current from another right moving hole, which would have further additional energy, 
so that energy conservation would be violated.

When the hole exits the wall, it has to become a particle again,
so that two particles are taken away from the right side of the wall.
Overall, the wall moves by {\em two} sites to the left in this quantum mechanical process,
contrary to the classical situation.

The particle-hole creation resembles Klein tunneling  \citen{KleinTunneling}.
However, here we have no external potential, but instead a {\em many-body} effect. 
In contrast to Klein tunneling, the dispersion is not linear, but a cosine. 
When one starts with an initially localized incoming particle \citen{supplement},
all momenta contribute, 
yet particle-hole transmutation
and the overall features of the transmission process are the same.

\begin{figure}[t!]
 \includegraphics[width=\columnwidth]{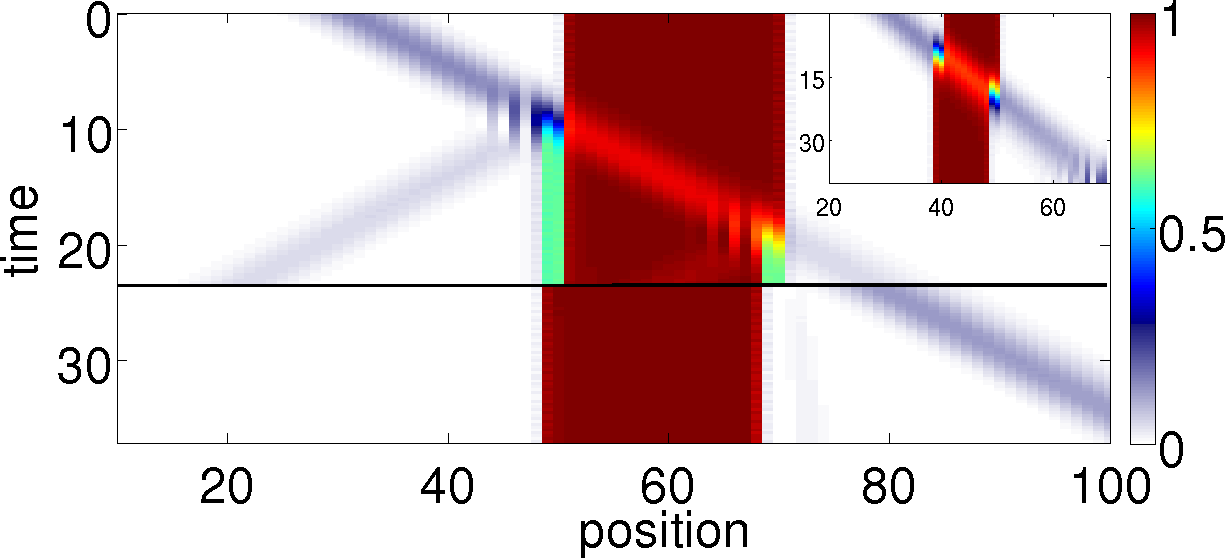}
\caption{
  Density of spin-polarized fermions with next nearest neighbor interaction.  
  Main figure: Nonintegrable model with $V=10$, $V_2 = 1.2$ and a 10 site wall.  
  Bottom part: density on condition that a particle is present to the right of the wall.
  Inset: Integrable model eq.\ref{eq:alpha} with $V=10, \alpha=0.1$ and same wall width.
}\label{fig:XXZNNN}
\end{figure}

\myheadingA{Small couplings; narrow walls} 
%
%
Particle-hole transmutation and the shift by two sites 
in the $tV$ model
are dictated by conservation laws
and the Pauli principle, 
{\em not} by especially strong coupling. 
Indeed, a coupling of e.g.\ $V/t=2.2$ ($\Delta=1.1$) still exhibits the same effects.
At such smaller $V/t$, M-string eigenstates are spatially more extended.
An initially prepared wall of $N$ consecutive particles evaporates more than at larger $V/t$ and becomes wider.
Yet \Fig{fig:smallDelta} shows that lack of backscattering and shift of the wall by 2 sites still occur 
at $V/t=2.2$.

Amazingly, a wall of only $N=2$ sites already shows the same phenomena, including a shift by two sites,
i.e.\ by the full wall thickness (\Fig{fig:ThreeModels}a, inset).

\myheadingA{Role of integrability} %
%
Conservation laws are essential for the observed effect.
We probe the role of integrability by studying a nonintegrable model
$  H_{V_2} = H_{tV} + V_2 \sum_i n_i n_{i+2}$ 
(\Fig{fig:XXZNNN}),
in which the energy current is not conserved.
Now there is indeed backscattering.
One might suspect the presence of a next-nearest-neighbor coupling to be responsible.
However, when one takes $H_{tV}$ and adds the conserved thermal current $J^E$ with next-nearest-neighbor terms,
\begin{equation}\label{eq:alpha}
H_\alpha = H_{tV} + \alpha j^E \,
\end{equation}
 another integrable model results, 
which does {\em not} show backscattering (\Fig{fig:XXZNNN}, inset).
This provides strong evidence that integrability is indeed closely connected to the observed lack of backscattering.

It remains an open question whether conservation of $n, E, j^E$ 
and the restricted local Hilbert space
are sufficient to suppress backscattering,
or whether full integrability is necessary.
It would be very interesting to study a nonintegrable model which conserves the above quantities,
if such a model exists\citen{Integrability-from-Q_three}.

\begin{figure}[t]
 \includegraphics[width=\columnwidth]{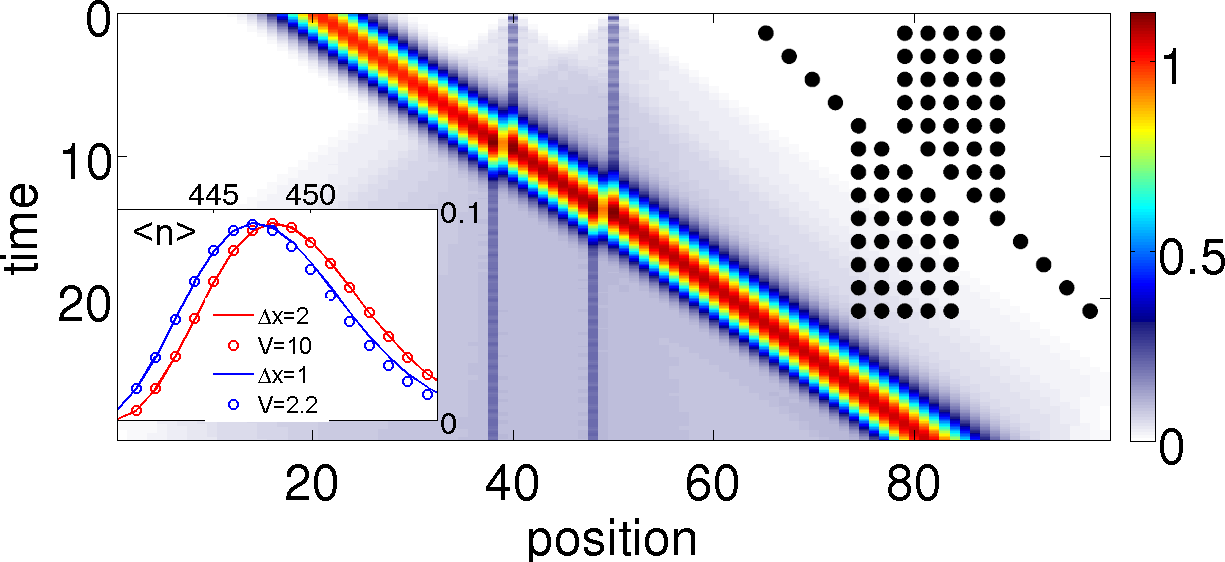}
\caption{Bipartite entanglement entropy $S_{AB}$ for the scattering in \Fig{fig:ThreeModels}a. 
        Right inset: semiclassical picture of the time evolution of particles, 
                     demonstrating both left-shift of the wall and right-shift of the travelling particle. 
                     Space and time directions are as in the main figure.
        Left inset:  spatial density distribution at large $x$; 
                     without wall (shifted right by 2 sites), and after passage through a wall.
}\label{fig:entropy}
\end{figure}

\myheadingA{Quantum mechanical nature of final state} 

In the bottom part of \Fig{fig:ThreeModels}(c) and \Fig{fig:XXZNNN}, the wave function has been projected 
(and then normalized) onto Fock states in which exactly one particle is 
present to the right of the wall, 
i.e. onto the case that the incoming particle was transmitted.
Then no reflection is visible in the reflected component
and we see that now the wall is shifted by 1 (resp. 2) site. 

Further insight is gained from the bipartite entanglement entropy \citen{Fazio-Review-Entanglement},
$  S_{AB}=-\textrm{tr}(\rho_{A}\log{\rho}_{A})$,
where 
$  \rho_A=\textrm{tr}_B \rho$, and $\rho=\ket{\psi}\bra{\psi}$ is the total density matrix.
$S_{AB}$ quantifies the number and strength of linear superpositions between $A$ and $B$.
When it 
is zero, then $\ket{\psi}$ is a product state $\ket{\psi}_A \ket{\psi}_B$.
In \Fig{fig:entropy} we show $S_{AB}$ as a function of time and of the position of the subsystem cut.
It is dominated by the entanglement inside the travelling Gaussian particle.
Additionally, the slight evaporation of the wall 
is visible as light blue cones emanating from the wall boundaries.
Strikingly, the amount of entanglement between the transmitted particle and the wall is 
hardly larger than on the left hand side of the wall 
\citen{footnote-entanglement}.
Thus there is no, or very little, entanglement due to the outgoing particle.
The outgoing particle, itself in an entangled Gaussian state,
is therefore to good precision
in a {product state} with the {\em shifted} wall.

One can understand further details from a semiclassical picture (\Fig{fig:entropy}, inset)
in which the incoming particle is thought of as a single occupied site.
Because of energy conservation, the closest that the particle can come to the wall is to a distance of one site;
then a particle from inside the wall has to move to the {\em left}, effectively propagating a hole inside the wall to the right.
This picture implies that the propagating signal should experience a shift forward by 1 lattice site 
both upon entry and upon exit of the wall, thus overall the transmitted particle should be shifted by 
{\em two sites} in forward direction.
Such a shift is indeed visible in the entropy in \Fig{fig:entropy}
and in the energy current \citen{supplement}.
The left inset in \Fig{fig:entropy}
shows that at large $V/t$, the Gaussian signal is moved forward by exactly 2 sites
{without noticeable change of shape}.
At small $V/t$, we observe a smaller shift of about $1.3$ sites, 
and we note that there will have been non-negligible additional scattering events with evaporated particles, 
which effectively widen the signal.

\begin{figure}[t]
 \includegraphics[width=\columnwidth]{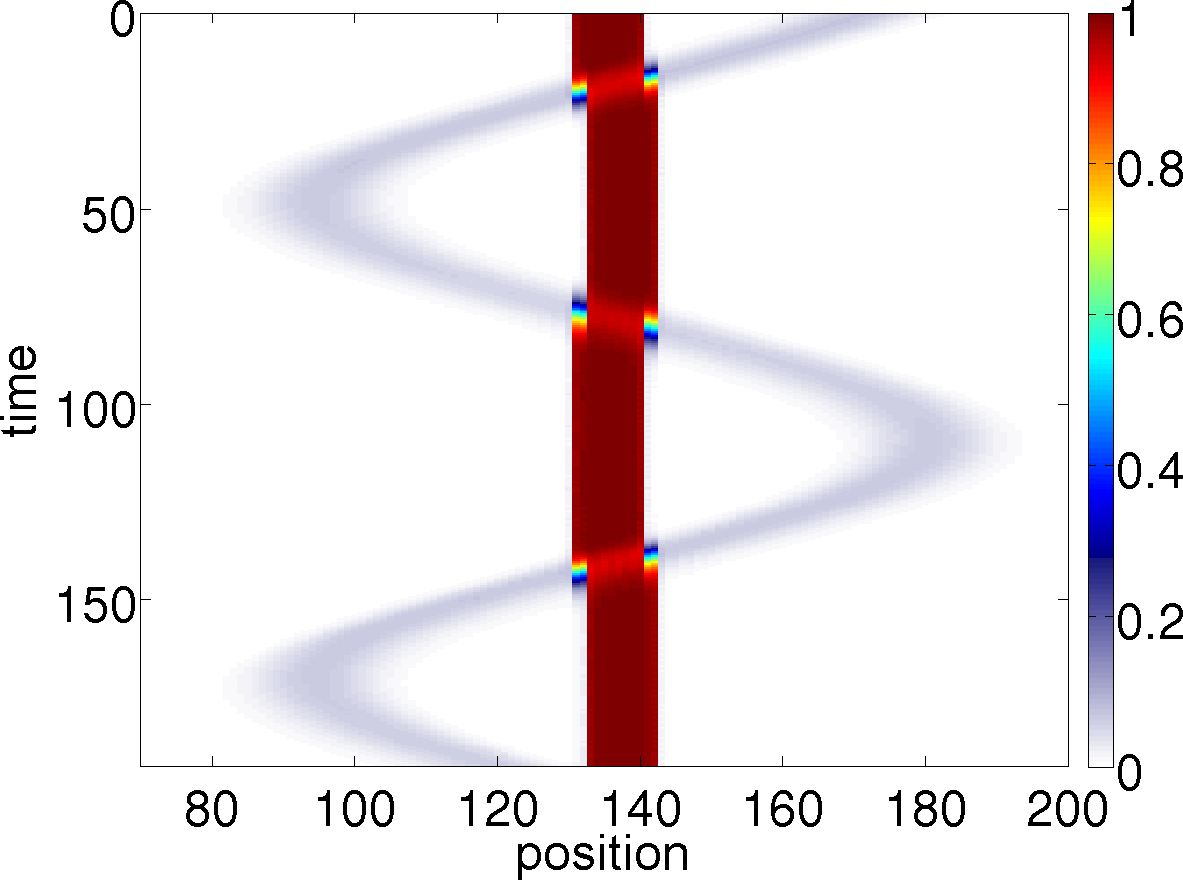}
\caption{A fermionic Quantum Newton's Cradle: density of spin polarized fermions in a linear electric field at $V=10$, 
         performing Bloch oscillations. 
         Initial Gaussian particle. $V=10$, 10-site wall.
}\label{fig:cradle}
\end{figure}

\myheading{Fermi Hubbard model}
%
The 1D fermionic Hubbard model is specified by 
\begin{equation}\label{eq:FH}
  H_{fH} = - t\sum_{i\sigma} \left( c_{i\sigma}^{\dagger}c_{(i+1)\sigma}+h.c.\right)  + U\sum_{i}
  n_{i\uparrow}n_{i\downarrow}
\end{equation}
where $U$ is the onsite interaction and $\sigma$ labels spin.
It is widely used as a basic model of strongly correlated matter. 
Its coherent dynamics have recently started to become accessible in 
cold atom experiments \citen{Schneider12}.

The initial state in \Fig{fig:ThreeModels}(b) contains $N=10$ consecutive doubly occupied sites.
Individually, they are repulsively or attractively bound by large $|U|$. 
They are however not mutually bound to each other. 
The outermost sites can therefore decay more easily than in model (\ref{eq:tV}),
and  $|U|\gtrsim 30$ is needed to clearly see the shift over background \citen{supplement}.

The model is integrable and
\Fig{fig:ThreeModels} shows no backscattering at all. 
There is a conserved current \citen{Takahashi-book}
(slightly different from the energy current). 
Again, since the wall is already doubly occupied, only a hole can move through.
Due to  energy conservation, the number of doubly occupied sites has to be conserved. 
Therefore a hole of {\em opposite} spin has to move.
Since there is no direct nearest neighbor interaction, 
in a semiclassical picture
the impinging particle can move up to the wall.
Therefore, unlike the spinless fermion case, there is no forward jump.
Indeed, we observe that transmission through the wall affects neither the shape nor the position 
of the wave packet \citen{supplement}.

\myheading{Bose Hubbard model}

The dynamics of the Bose Hubbard model
\begin{equation}
H_{bH} =  -t\sum_{i} \left( b_i^{\dagger}b_{i+1} + h.c. \right)  + U\sum_{i}
n_i\left({n_i-1}\right)  \label{eq:bosehub} .
\end{equation} 
is now widely realized in experiments with ultra cold atoms in optical lattices \citen{Cold-atom-review}.
Because of Bose statistics, 
there is no restriction on local occupation numbers.

The initial state in \Fig{fig:ThreeModels}(c) contains $N=10$ consecutive sites with two bosons each, 
which are attractively or repulsively bound on each site.
For the Bose-Hubbard model, the stability of doubly occupied walls grows with the wall width, 
as well as with $|U|$,  
because the pairs can bind to each other when they are on 
neighboring sites \citen{Bose-Hubbard-double-dimer-stability}.

The Bose-Hubbard model is non-integrable. The scattering result 
is similar to \Fig{fig:XXZNNN}, with partial backscattering.
Because of energy conservation, 
any transmitted incoming signal
has to go through the wall as a hole, 
leaving behind an additional particle and thus one additional doubly occupied single site to the left of the wall.
The projected part at the bottom of \Fig{fig:ThreeModels}c shows that for these transmitted particles,
the wall is indeed again shifted by 1 site,
which remains visible for $|U|\gtrsim 10$. 
Note that inside the wall, the velocity of the hole is twice as high as outside, 
due to the double occupancy inside the cluster which renormalizes the hopping.

\begin{figure}[t]
 \includegraphics[width=\columnwidth]{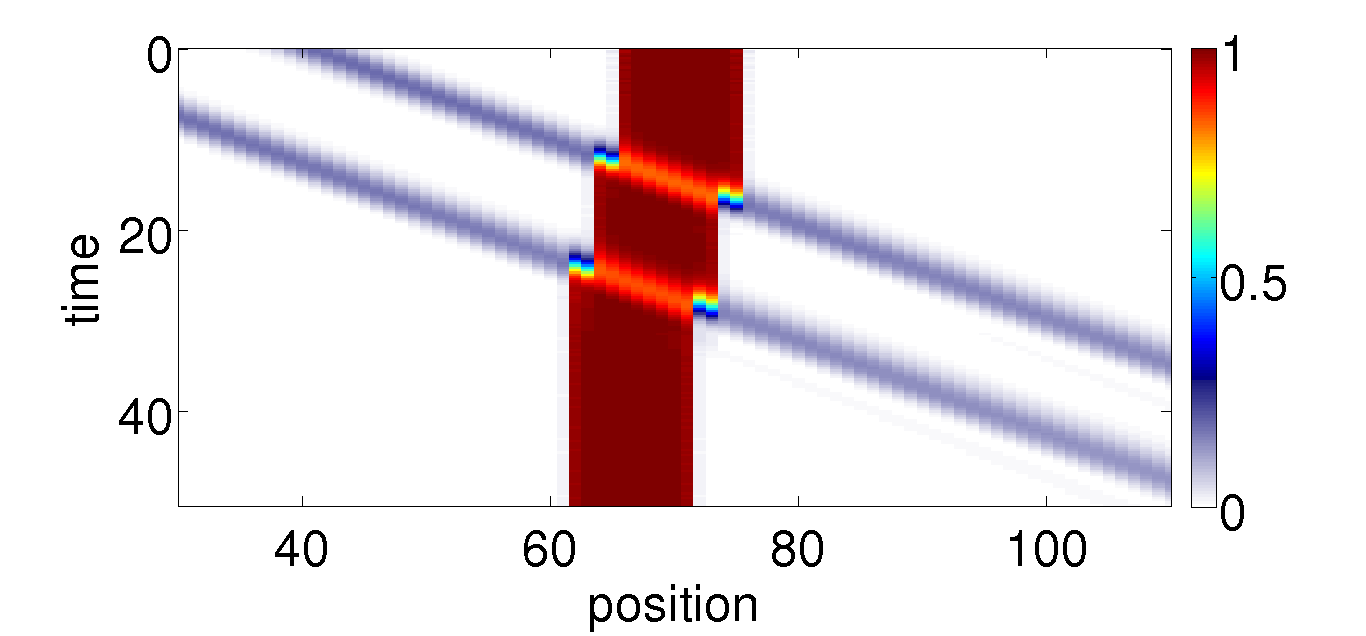}
\caption{Atomic scale signal counter and shift register: each passing particle shifts the wall by 2 sites. 
        (Density of spin-polarized fermions, $V=10$, 10-site wall.)
}\label{fig:shiftregister}
\end{figure}

\myheading{Applications} 
%
We discuss some immediate applications, making use of the wall shifts,
the clean nature of scattering in the integrable $tV$ model of spin-polarized fermions
(or spin $1/2$ Heisenberg chain), and the large stability of walls.

\myheadingB{Fermionic quantum Newton's cradle on a lattice.}
The continuous space ``Quantum Newton's Cradle'' \citen{Kinoshita-cradle}  
is one of the most famous experiments with cold atoms.
Here we construct a rather distinct lattice fermionic version 
by placing the system into an electric field with constant gradient, 
adding $\sum_j \,0.06\, j\, \hat n_j$ to eq.~(\ref{eq:tV}).
An incoming Gaussian particle then experiences Bloch oscillations \citen{Bloch-oscillations},
whereas the wall is not affected noticeably due to its high mass.
\Fig{fig:cradle} shows the result: a periodic motion very similar to the classical Newton's cradle,
except that at each impact, the wall moves by {\em two} sites instead of one.

\myheadingB{Qubits and atomic scale shift register.}
When several individual particles hit a wall in succession, the shifts 
add up quasi-classically, as shown in \Fig{fig:shiftregister}.
In effect, the wall position counts the incoming particles,
of potential practical interest, e.g.\ in spintronics applications.
Furthermore, when a bound pair of particles impinges on the wall (not shown),
it is transmitted inside the wall as a hole pair, shifting the wall by 4 sites after transmission.

One way to encode a qubit with M-strings is to assign $|0\rangle$ to an M-string in a certain position
and $|1\rangle$ to a similar M-string in a different position, ideally non-overlapping.
We note that for large values of $V/t$ and $M$, this qubit will decay only on timescales exponentiall small versus the inverse hopping $1/t$.

A sequence of walls and empty space, possibly of various thicknesses, 
can be interpreted as a sequence of bits.
A suitable quantum-mechanical superposition of such sequences can be seen as a sequence of {qubits}.
An incoming single particle would 
shift the complete sequence of qubits coherently by 2 sites, without becoming entangled,
making for a coherent qubit shift register \citen{atomic-shift-register}.

\begin{figure}[t]
 \includegraphics[width=\columnwidth]{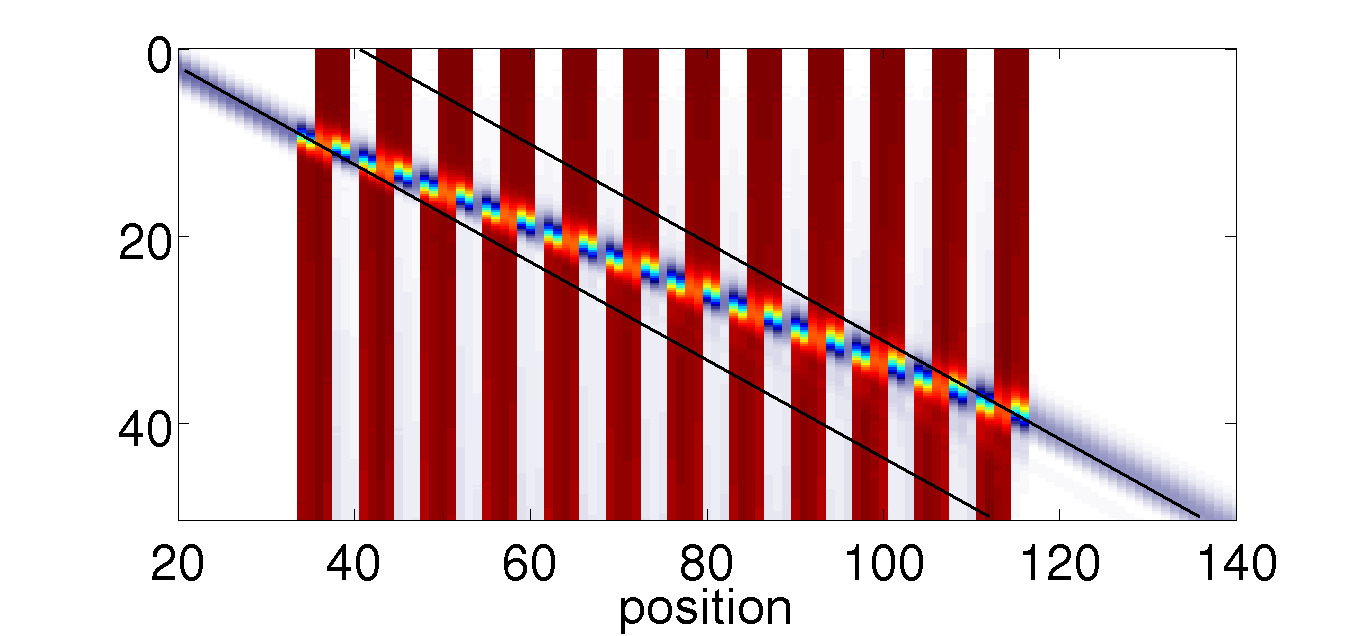}
\caption{Metamaterial with ``tachyonic'' mode: transition of a spin-polarized fermion through a set of 12 equally spaced 4-particle clusters 
         with an intra-cluster distance of 3 sites, at $V/t$=20. 
         Particle density shown.
         The forward shift of the travelling particle at each wall leads to an average velocity 
         larger than the velocity 
         of a particle on an empty lattice.
}\label{fig:manywalls}
\end{figure}

\myheadingB{Metamaterial, tachyon.}
We have shown above that the transmitted particle is shifted { forward} by 2 sites (\Fig{fig:entropy}, inset).
When the particle moves through several walls in succession, as shown in \Fig{fig:manywalls},
the individual shifts of the travelling particle add up. 
It moves with an average velocity {\em higher} than on the empty lattice.
If one regards the latter as a "vacuum" with velocity $v_0$, 
then the sequential walls act like a metamaterial with a tachyonic mode
of velocity $v_0 d/(d-2)$, where $d$ is the wall spacing ($d=7$ in \Fig{fig:manywalls}).
The results in \Fig{fig:manywalls} match this expectation precisely.

\myheading{Conclusions}
We have shown that the quantum mechanical transmission of a particle through a wall of neighboring particles
exhibits surprising effects, namely pair creation with particle-hole transmutation and 
a shift of the wall.
In the spin-polarized fermion or Heisenberg case, the wall shifts by {\em two} sites,
and the transmitted particle jumps forward by two sites at large couplings.
In addition, and independently, we find that there is no backscattering in the integrable models studied.
These effects are due to conservation laws and the discrete nature of particles.
They are therefore robust and, e.g., still occur at small anisotropies $\Delta \gsim 1$ in the Heisenberg model
and for very narrow walls. 
The final state is close to a product state of a shifted wall and a transmitted particle of unchanged shape.
Applications for spintronics may be possible, like an atomic scale signal counter or coherent shift of qubits.
Last, but not least, the phenomena discussed should come within reach experimentally
with cold atoms in optical lattices in the foreseeable future.
\\

\textit{Acknowledgements.} 
We would like to thank I. Bloch, M. Cheneau, J.S. Caux, J. Mossel, Chr.~Gro\ss, F. Heidrich-Meisner, M. Knap, 
U. Schneider, V. Korepin, and especially F. Essler
for valuable comments and discussions. 
This work was supported by the Austrian Science Fund (FWF) within the SFB ViCoM (F41). 
MH and HGE thank the KITP for hospitality. This research was supported in part by the 
NSF under grant No. NSF PHY05-51164.



\setcounter{figure}{0}    \renewcommand{\thefigure}{S\arabic{figure}}
\setcounter{equation}{0}  \renewcommand{\theequation}{S.\arabic{equation}}  
\setcounter{page}{1}      \renewcommand{\thepage}{S.\arabic{page}}  


\mytitle{Supplementary Material 
     \\[1ex]     Quantum Bowling: Particle-Hole transmutation in one-dimensional strongly interacting lattice models
  }
\myauthor{Martin Ganahl}
\myaffiliation{Institut f. Theoretische Physik, Technische Universit\"at Graz, Petersgasse 16, 8010 Graz, Austria}
\myauthor{Masud Haque}
\myaffiliation{Max-Planck-Institut f\"ur Physik komplexer Systeme, N\"othnitzer Stra\ss e 38, 01187 Dresden, Germany}
\myauthor{H.G. Evertz}
\email{evertz@tugraz.at}
\myaffiliation{Institut f. Theoretische Physik, Technische Universit\"at Graz, Petersgasse 16, 8010 Graz, Austria}
\date{} 

\fancyhead[C]{Supplement}

\clearpage 
\appendix
\section{\large Supplementary material}


      \begin{figure}[b]
          \includegraphics[width=\columnwidth]{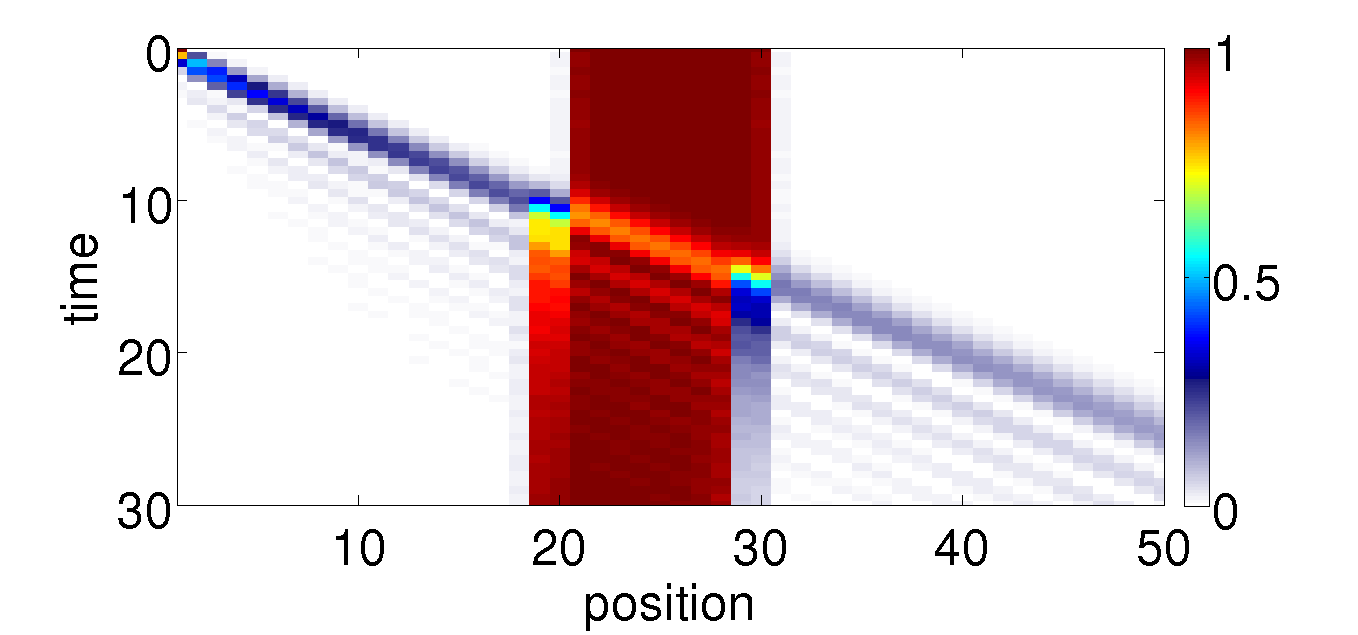}
   \caption{Particle density for a single magnon-like excitation
            hitting a wall of $N=10$ sites in the spin-polarized fermion model, $V=10$.}
        \label{fig:XXZ_magnon}
      \end{figure}

      \begin{figure}[b]
  \includegraphics[width=\columnwidth]{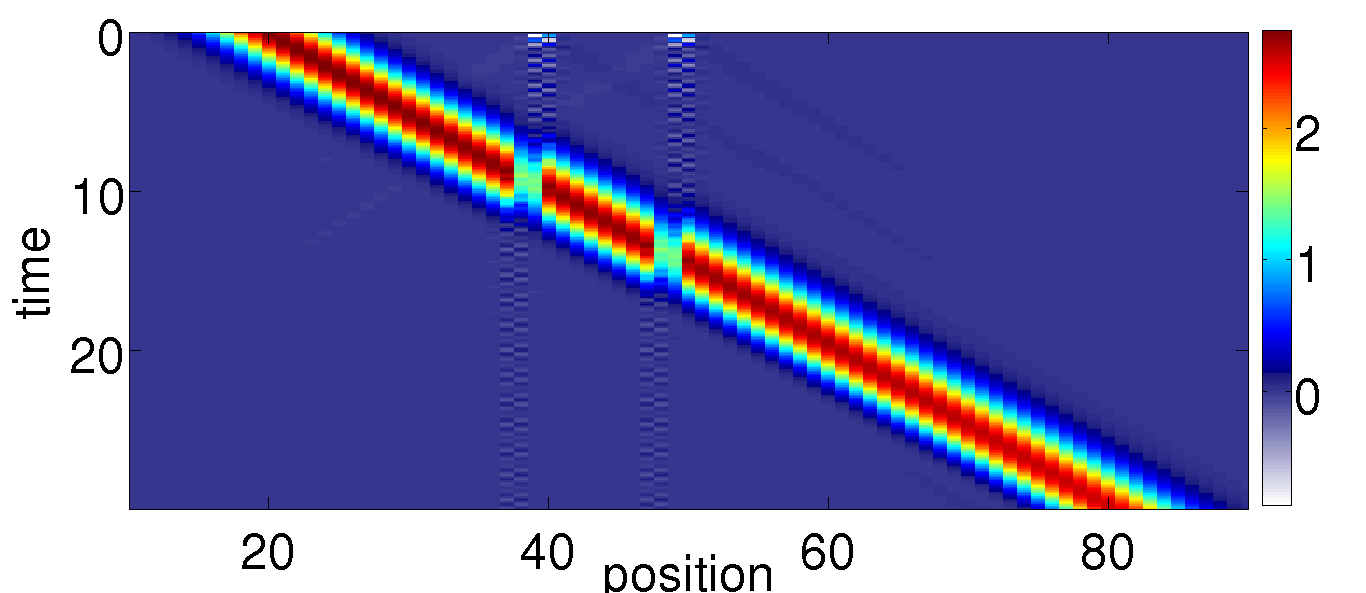}
   \caption{Energy current for a Gaussian particle hitting a wall of $N=10$ sites (parameters of \Fig{fig:ThreeModels}a).}

        \label{fig:XXZ_current}
      \end{figure}

 \subsection{Methods}
 

We calculated the full quantum manybody time evolution of all models by
time evolved block decimation (TEBD) \citen{s-tebd},
which is closely related to the time dependent Density Matrix Renormalization Group (tDMRG) \citen{s-tDMRG}.
These techniques are based on Matrix Product State (MPS) 
representations and allow for high precision \citen{s-SchollwoeckReview},
especially with the relatively low entanglement in our simulations.
We made use of local particle number conservation and employed matrix dimensions of 200 -- 400.
We verified results against full diagonalization, where applicable. 
For the Bose-Hubbard model, we limited the local occupation number on each site to $n_{max}=5$,
which was high enough to ensure that the results did not depend on $n_{max}$.

  \subsubsection{Initial state.}
We started with an empty chain (length as in the figures) with open boundary conditions.
On this lattice we put a product state of $N$ consecutive singly or doubly filled sites to make the wall,
by specifying MPS matrices accordingly.
In addition, we started an incoming particle, either in the shape of a Gaussian excitation, or of a localized magnon,
as described below.

  \subsubsection{Creation of Gaussian excitations.} 
We followed ref.~\citen{s-ulbricht}
and applied 
\begin{align}
 &\sum_x\exp(-\frac{(x-x_0)^2}{2\sigma^2})\exp(i (x-x_0)k_0)\, c_{x}^{\dagger}\nonumber\\
  \sim  &\int dk \exp(-\frac{(k-k_0)^2\sigma^2}{2})\exp(-ikx_0)\, c_k^{\dagger}
\end{align}
to the state with the initial wall,
with momentum $k_0=-\pi/2$, width $\sigma= 4$, 
and center position $x_0$ as visible in the figures,
and then normalized to create a single Gaussian shaped particle. 
Its velocity is
$\frac{d\epsilon}{dk}|_{k_0}$, 
where   $\epsilon(k)=-2t\cos(k)$
is the single particle dispersion.
With $k_0=-\pi/2$, the resulting 
particle travels at velocity $2t$ with almost no dispersion,
as seen in the figures.

  \subsubsection{Single magnon excitation}
Alternatively, we started a ``magnon'' excitation by adding a single particle at the first left-hand site of the lattice.
This procedure may be easier to implement experimentally.
We also used it in the Bose-Hubbard model, because of better visibility  of the transmitted hole  inside the wall
than with a Gaussian particle.

The behavior of such an excitation is easiest to understand by considering an empty lattice
with a single particle in the middle at a site $x=0$ at time $t=0$ \citen{s-Ganahl11}.
Since a single particle cannot interact, all models considered in this paper are equivalent to
tight binding fermions in this case.
The initial state is $\ket{\psi(0)} = \ket{1}_{x=0} = \sum_k \ket{1}_k$.
Then $\ket{\psi(t)} = \exp(-iHt)\ket{\psi(0)} 
                    = \exp(i 2t \sum_p \cos p \hat n_p) \sum_k \ket{1}_k
                    = \sum_k \exp(i 2t \cos k) \ket{1}_k 
$,
which can be written as  a Bessel function \citen{s-Ganahl11}.
Each $k$ mode travels with a velocity $v_k=-2t\frac{d\cos k}{dk}= 2t\sin k \le 2t$.
Close to the maximum velocity $v_{max}=2t$, the most modes contribute,
which produces a magnon, a linearly propagating wave distinctly visible in space-time figures.
Modes which are further away from $k=\pi/2$ produce additional oscillatory behavior.
When starting in the middle of an empty lattice, a left-moving and a right-moving branch ensue,
each with an overall probability of $1/2$ of containing the particle.
When started close to an open boundary, the particle travels only in one direction, away from the boundary,
with unit probability.
We note that the distinctly visible fastest branch has a velocity $v_{max}=2t$ at its fastest edge,
whereas the location of maximum intensity slowly moves away from the edge sublinearly. 
After finite times, the average velocity of the peak is therefore slightly smaller than $v_{max}$,
but it converges towards $v_{max}$ for large times.

      \begin{figure}[tb]
          \includegraphics[width=\columnwidth]{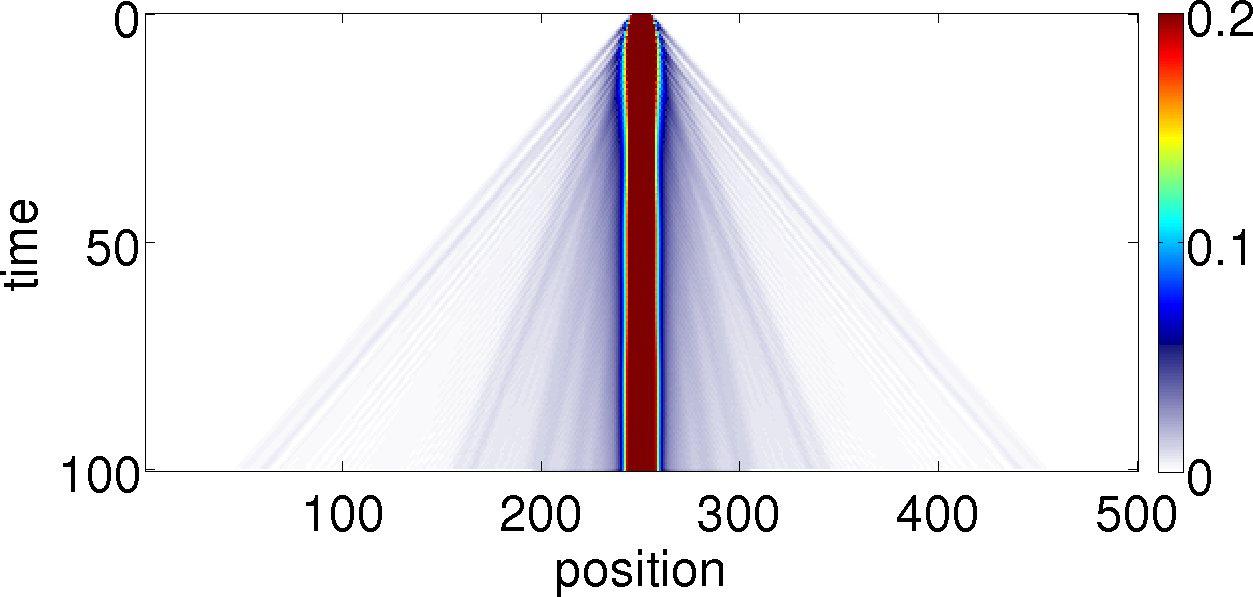}
          \caption{Particle density for evaporative cooling of a 10-site wall at $V=2.2$. 
                   The vertical scale has been cut off at $n=0.2$ for better visibility.}
        \label{fig:XXZ_Cooling}
      \end{figure}

      \begin{figure}[tb]
          \includegraphics[width=\columnwidth]{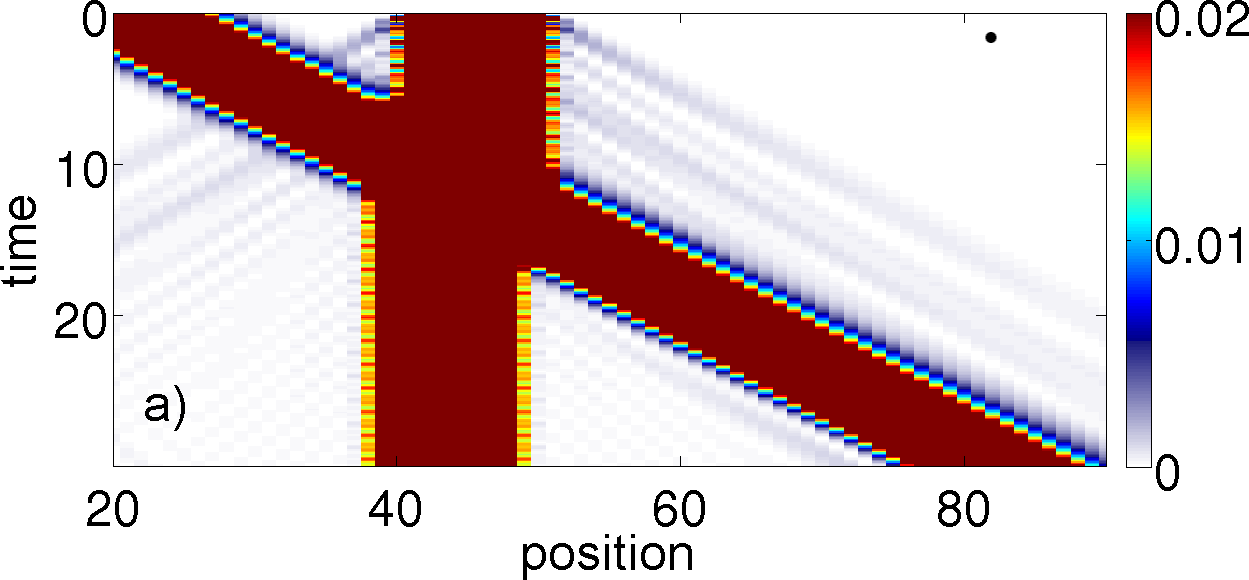}
   \caption{Particle density at $V=10$. Same as \Fig{fig:ThreeModels}a in the main paper,
            but with vertical scale cut off at very low $n=0.02$ to make evaporation of wall visible.}
        \label{fig:XXZ_evaporation}
      \end{figure}

In \Fig{fig:XXZ_magnon} 
we show the analogue of \Fig{fig:ThreeModels}a in the main text, with an initial magnon
impinging on the wall, instead of a Gaussian particle.
We see that the total intensity of the incoming particle now arrives over time in several waves.
Each of them behaves similar to \Fig{fig:ThreeModels}a. There is no backscattering,
and the probability distribution of the wall converges to a complete shift of 2 sites.

  \begin{figure}[tb]
      \includegraphics[width=\columnwidth]{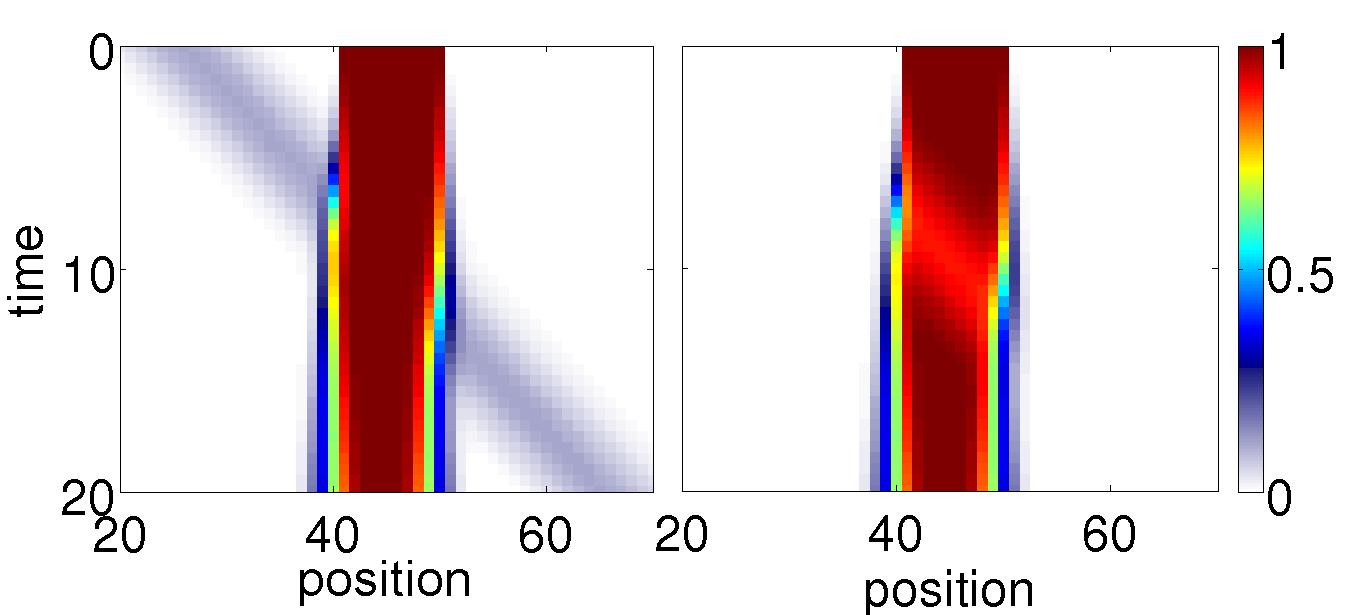}
    \caption{Densities in the Fermi Hubbard model at $U=30$, similar to \Fig{fig:ThreeModels}b in the main text.
             Left: up-spin, right: down-spin. The wall widens over time, 
             but a shift of 1 lattice site (2 particles) is visible in the spatial density distribution.   } 
        \label{fig:FH_U30}
  \end{figure}

  \begin{figure}[tb]
      \includegraphics[width=\columnwidth]{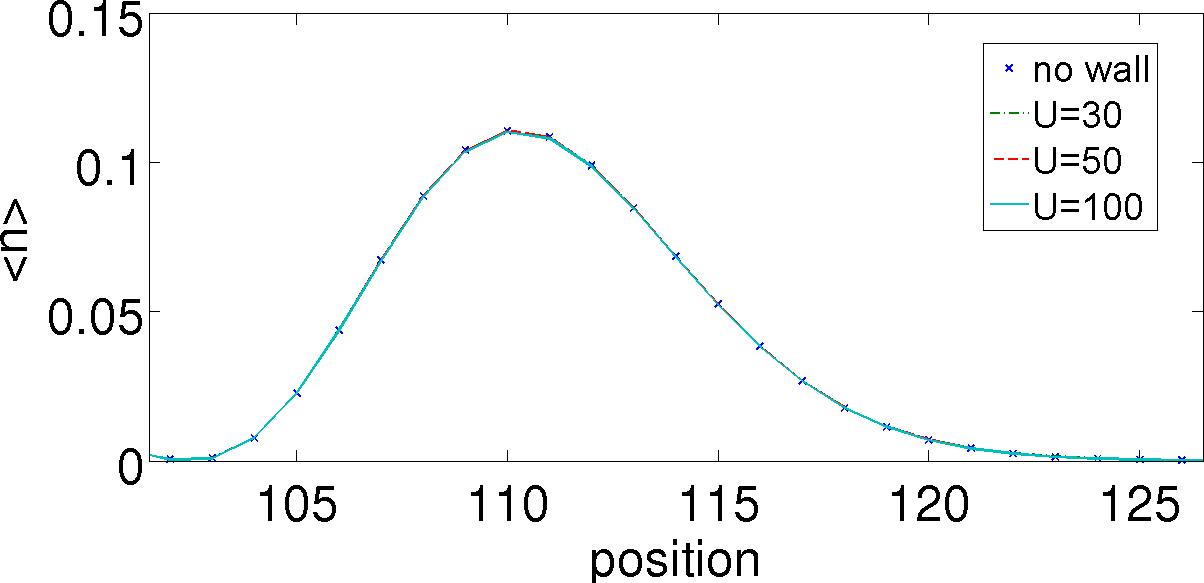}
    \caption{Fermi Hubbard model. The spatial density distribution of the transmitted particle 
             is independent of $U$,
             and independent of whether there is a wall or not. }
        \label{fig:FH_shape}
  \end{figure}

 \subsection{Energy current for spin polarized fermions}
\Fig{fig:XXZ_current} shows the energy current corresponding to \Fig{fig:ThreeModels}a in the main text.
The current is conserved globally, and in spatially disconnected regions also locally.
The local nature of the incoming Gaussian particle and its jump forward at both edges of the wall are clearly visible.

  \subsection{Evaporation and evaporative cooling.}
At $V/t = 2.2$ we prepared a wall of $N=10$ particles and let it evolve for time $100/t$,
during which particles evaporated and the wall became wider, closer to a 
linear combination of M-string eigenstates with mostly large M
\citen{s-MosselCaux}, in a kind of  evaporative cooling.
\Fig{fig:XXZ_Cooling} clearly shows the evaporation of single particles and of slower $M=2$ 
bound strings.
We then cut the remaining state to a width of 110 sites 
and started a Gaussian particle, as visible
in \Fig{fig:smallDelta} in the main paper.

For comparison, \Fig{fig:XXZ_evaporation} shows the much smaller scale of evaporation at $V/t=10$ for the time evolution
in \Fig{fig:ThreeModels}a of the main paper.

  \subsection{Fermi Hubbard model}  

\Fig{fig:FH_U30} displays scattering for the Fermi Hubbard model, similar to 
\Fig{fig:ThreeModels}b of the main paper, but at smaller coupling $U/t=30$.
The wall decays faster in this case. Its shift is still visible in the spatial density distribution
before and after the scattering.
\Fig{fig:FH_shape} shows that, within the precision of our data, 
the shape of the transmitted particle is independent of $U$.
It is not influenced by additional scattering events with evaporated particles.
This indicates that the particle emerges practically unchanged, 
and without phase shift, from each scattering for a wide range of $U$ values.

\subsection{Interaction inversion symmetry}

In the appendix of ref.~\onlinecite{s-Schneider12}
it was shown for the fermionic Hubbard model that the time evolution is invariant under the transformation $U \rightarrow -U$,
for observables invariant under a combined transformation of time reversal and $\pi$-boost $c_q \rightarrow c_{q+Q}$, which include particle density,
and for initial states which only aquire a phase under the combined transformation.
We note that the proof applies to density-density interactions in general, including the $tV$-model and the bosonic Hubbard model,
and to initial product states. For an initial state including a Gaussian, it applies when $k_0 \rightarrow -k_0$ is also transformed.
All our results therefore apply to both attractive and repulsive models, i.e.\ they are invariant under $U\rightarrow -U$, resp.\ $V\rightarrow -V$.




\begin{thebibliography}{1}
\bibitem{footnote-entanglement} 
 There is a difference of up to 0.007, which may be caused by earlier scattering with evaporated particles.
\end{thebibliography}
\end{document}